\documentclass[aps,prl,twocolumn,amsmath,amssymb,nofootinbib,superscriptaddress]{revtex4}

\newcommand{\bra}[1]{\langle#1|}
\newcommand{\ket}[1]{|#1\rangle}

\usepackage[pdftex]{graphicx}
\usepackage{mathrsfs}
\usepackage[colorlinks]{hyperref}

\begin{document}

\bibliographystyle{apsrev}

\title{Sampling arbitrary photon-added or photon-subtracted squeezed states is in the same complexity class as boson sampling}

\author{Jonathan P. Olson}
\email[]{jolson7@lsu.edu}
\affiliation{Hearne Institute for Theoretical Physics and Department of Physics \& Astronomy, Louisiana State University, Baton Rouge, LA 70803}

\author{Kaushik P. Seshadreesan}
\email[]{ksesha1@lsu.edu}
\affiliation{Hearne Institute for Theoretical Physics and Department of Physics \& Astronomy, Louisiana State University, Baton Rouge, LA 70803}

\author{Keith R. Motes}
\affiliation{Centre for Engineered Quantum Systems, Department of Physics and Astronomy, Macquarie University, Sydney NSW 2113, Australia}

\author{Peter P. Rohde}
\email[]{dr.rohde@gmail.com}
\homepage{http://www.peterrohde.org}
\affiliation{Centre for Engineered Quantum Systems, Department of Physics and Astronomy, Macquarie University, Sydney NSW 2113, Australia}
\affiliation{Centre for Quantum Computation and Intelligent Systems (QCIS), Faculty of Engineering \& Information Technology, University of Technology, Sydney, NSW 2007, Australia}

\author{Jonathan P. Dowling}
\affiliation{Hearne Institute for Theoretical Physics and Department of Physics \& Astronomy, Louisiana State University, Baton Rouge, LA 70803}
\affiliation{Computational Science Research Center, Beijing 100084, China}

\date{\today}

\frenchspacing

\begin{abstract}
Boson sampling is a simple model for non-universal linear optics quantum computing using far fewer physical resources than universal schemes. An input state comprising vacuum and single photon states is fed through a Haar-random linear optics network and sampled at the output using coincidence photodetection. This problem is strongly believed to be classically hard to simulate. We show that an analogous procedure implements the same problem, using photon-added or -subtracted squeezed vacuum states (with arbitrary squeezing), where sampling at the output is performed via parity measurements. The equivalence is exact and independent of the squeezing parameter, and hence provides an entire class of new quantum states of light in the same complexity class as boson sampling.
\end{abstract}

\maketitle
\section{Introduction}
Scalable quantum computing \cite{bib:NielsenChuang00} is likely to usher in a new age for computing. Certain problems, such as integer factorization \cite{bib:Shor97}, search algorithms \cite{bib:Grover96} and quantum simulation \cite{bib:LloydSim} are believed to be more efficient on quantum computers than on classical computers. Whilst there are a number of differing models for realizing scalable quantum computing, linear optics quantum computing (LOQC) \cite{bib:KLM01, bib:KokLovett11} appears to be one of the most promising.  Photons are not only relatively easy to prepare, manipulate, and measure, but also also have very long decoherence times.  Unfortunately, the hurdles for implementing full universal LOQC remain very challenging and appear to be impractical with current technologies. Hence, there is much interest in simpler, more feasible approaches that could be demonstrated with existing technology.

In this spirit, Aaronson \& Arkhipov (AA) introduced the boson sampling model \cite{bib:AaronsonArkhipov10, bib:Chapter}. Whilst not universal for quantum computing, boson sampling uses only passive linear optical elements to efficiently implement a particular sampling problem, which is strongly believed to be hard on a classical computer. This makes boson sampling vastly simpler than full-fledged LOQC because it does away with some of the more challenging experimental requirements, namely fast-feedforward, optical quantum memory, and the need for a plethora of optical elements.

The mere fact that boson sampling implements a computationally hard problem using technologies that are, for the larger part, available today makes it of great practical interest. Its relative simplicity and frugal physical resource requirements may render it the route towards building the first post-classical quantum computer. Recently, there have been numerous elementary experimental demonstrations of boson sampling using three photons \cite{bib:Broome2012, bib:Spring2, bib:Crespi3, bib:Tillmann4}. Also, there have been proposals for scalable implementations of boson sampling in optical systems and ion traps~\cite{bib:Motes13, bib:Lund13, MGDR14, Shen_14}. As a first application of the model, boson sampling (with a suitably modified input state) has been shown to yield a practical tool for difficult molecular computations to generate molecular vibronic spectra~\cite{HGPMA14}.

Recent research efforts include showing certification of true boson sampling to distinguish it from uniform sampling, classical sampling, or random-state sampling~\cite{bib:AA13response, Spagnolo_13, Carolan_13, bib:Molmer13, QKUM14}. The impact of mode-mismatch, spectra of the bosons, and spectral sensitivities of detectors in realistic implementations of boson sampling have also been studied~\cite{Shc14a, Rho14}. This has further paved the way to a theory of interference with partially indistinguishable particles, where any realistic imperfections in the source and detectors can be completely characterized~\cite{Shc14b, Tich14}.

In other theoretical considerations, the surprising discovery of the complexity of sampling Fock states via linear optics opened inquiry into the complexity of other linear optical systems. The obvious open question is `are there other quantum states of light, other than Fock states, which also yield computationally hard sampling problems?' To this end, several other quantum states of light have been shown to implement likely-hard sampling problems similar to AA's original boson sampling. Gaussian states, when measured in a Gaussian basis, are known to be classically simulatable~\cite{bib:Bartlett02, bib:Bartlett02b}. Sampling Gaussian states in the photon-number basis, however, has attracted recent interest in light of boson sampling. It has been shown that sampling some Gaussian states with photon number counting can be just as hard as boson sampling~\cite{bib:Lund13}. More specifically, while thermal state inputs can be simulated efficiently by a classical algorithm~\cite{RLR_14}, sampling two-mode squeezed vacuum states can be hard to simulate~\cite{bib:PhysRevA.88.044301, bib:Lund13}. Photon-added coherent states have been shown to implement computationally hard sampling problems in the photon number basis in the low amplitude limit~\cite{bib:Pacs13}. Sampling generalized cat states (arbitrary superpositions of coherent states, \mbox{$\sum_i \lambda_i \ket{\alpha_i}$}) have also been considered~\cite{bib:RohdeCat} and shown to be computationally hard for sampling in the photon-number basis.

Here we will demonstrate that, in general, boson sampling using photon-added or -subtracted squeezed vacuum (PASSV) states and parity measurements yields a computational problem of equal complexity to Fock state boson sampling in \emph{all} parameter regimes. Importantly, because the mapping is exact, AA's robustness result for approximate boson sampling holds.  Note that experimental implementation of PASSV sampling is not the focus of our result, as doing so is more difficult than single photon boson sampling.  Our goal is to provide clarity on the theory of classifying the sampling complexity of quantum states.  In particular, we wish to demonstrate that Fock states are not unique -- on the contrary, there are a plethora of other quantum states of light which yield sampling problems with similar complexity to boson sampling. Nevertheless, we believe it is still important to show that such a device is still physically realizable.

\section{PASSV Sampling}
In order to show that the complexity of the boson sampling model introduced by AA also extends to PASSV sampling, we prove that it implements the same logical problem, i.e. that the output of the device corresponds to the same matrix permanent sampling problem as in AA boson sampling.  The advantage of this method is that it allows us to avoid the very lengthy analysis comprising AA's original complexity proof, yet we can still apply all of the same results.  However, one must be careful to show equivalence throughout the problem.  

Both models employ a similar general setup; $m$ optical input modes are fed into a passive, linear interferometer and the resulting output is measured in each mode, with the joint distribution of the measurement constituting one sample.  However, the details differ in each step (which we will classify by \textbf{input}, \textbf{evolution}, \textbf{output}, and \textbf{measurement}).  To carefully guide the reader, we will first provide the details of each step of both models head-to-head, discussing the relevant differences.  We will then proceed to show that the two models implement the same sampling problem, and thus exhibit the same computational complexity.  For consistency and simplicity, we will consider the case of photon-added states throughout the comparison.

\subsection{Contrast with Fock State Boson Sampling}
We now provide a detailed comparison for each step of the model.

\textbf{Input:} 
AA's Fock state boson sampling begins by preparing the first $n$ modes of a passive linear optics interferometer with single photons and the remaining \mbox{$m-n$} modes with vacuum states, where \mbox{$m=\Omega(n^2)$} (i.e. $m$ is asymptotically bounded below by some positive constant times $n^2$).  As conjectured by AA, this requirement ensures that the probability of more than one photon arriving at a given output mode is small (sometimes referred to as the `bosonic birthday paradox'). A stronger requirement of $m=\Omega(n^6)$ will suffice if one does not wish to adopt this additional conjecture. The input state is thus,
\begin{eqnarray} \label{eq:input_state}
\ket{\psi}_\mathrm{in}^\mathrm{AA} &=& \ket{1_1,\dots,1_n,0_{n+1},\dots,0_m} \nonumber \\
&=& \hat{a}_1^\dag\dots \hat{a}_n^\dag \ket{0_1,\dots,0_m},
\end{eqnarray}
where subscripts denote mode number and $\hat{a}_i^\dag$ is the photonic creation operator on the $i$th mode.

In contrast, for PASSV boson sampling we prepare the first $n$ modes of a similar interferometer with PASSV states and the remaining \mbox{$m-n$} modes with squeezed vacuum (SV) states. We let the squeezing parameter $\xi$ be arbitrary, but ensure each mode has the same amount of squeezing. In the case of photon-added states, the input state is thus,
\begin{eqnarray} \label{eq:input_state}
\ket{\psi}_\mathrm{in}^\mathrm{SV} &=& \hat{a}_1^\dag\hat{S}_1(\xi)\dots\hat{a}^\dag_n\hat{S}_n(\xi)\hat{S}_{n+1}(\xi)\dots \hat{S}_m(\xi)\ket{0_1,\dots,0_m} \nonumber \\
&=& \hat{a}^\dag_1\dots \hat{a}^\dag_n\ket{\xi_1,\dots,\xi_m},
\end{eqnarray}
where we have abbreviated $\hat{S}_i(\xi)\ket{0_i}=\ket{\xi_i}$ and again the subscript indicates mode number (not separate variables).  The state in Eq.(\ref{eq:input_state}) is not normalized, but this can be corrected by considering the state $\mathcal{N}\ket{\psi}_\mathrm{in}^\mathrm{SV}$ where
\begin{equation} \label{eq:normalization}
\mathcal{N}= \Big[\sqrt{1+\sinh^2(\xi)}\;\Big]^{-n}.
\end{equation}
Since the normalization does not affect our result, we leave it out of subsequent equations for simplicity.  Here,
\begin{equation}
\hat{S}(\xi)=\exp\left[\frac{1}{2}(\xi^*\hat{a}^2-\xi\hat{a}^\dag{}^2)\right],
\end{equation}
is the squeezing operator and $\hat{a}^\dag$ and $\hat{a}$ are the photon creation and annihilation operators respectively. In the Fock basis, if \mbox{$\xi=re^{i\theta}$}, then \mbox{$\hat{S}(\xi)\ket{0}=\ket{\xi}$} has the representation \cite{bib:GerryKnight04},
\begin{equation} \label{eq:sv}
\ket{\xi}=\frac{1}{\sqrt{\cosh(r)}}\sum_{m=0}^\infty (-1)^m \frac{\sqrt{(2m)!}}{{2^m m!}}e^{im\theta}\tanh^m(r)\ket{2m},
\end{equation}
and thus the SV state contains only even photon-number terms.  From the action of the creation or annihilation operator, a PASSV state then contains only odd photon-number terms. In the limit of vanishing squeezing, the SV state approaches the vacuum state, \mbox{$\lim_{\xi\to 0}\ket{\xi} = \ket{0}$}, and the photon-added SV state approaches the single-photon state, \mbox{$\lim_{\xi\to 0}\hat{a}^\dag\ket{\xi} = \ket{1}$}. Thus, we see that in the limit of vanishing squeezing, photon-added SV boson sampling reduces to ideal Fock state boson sampling.

\begin{figure}[!htb]
\includegraphics[width=0.4\columnwidth]{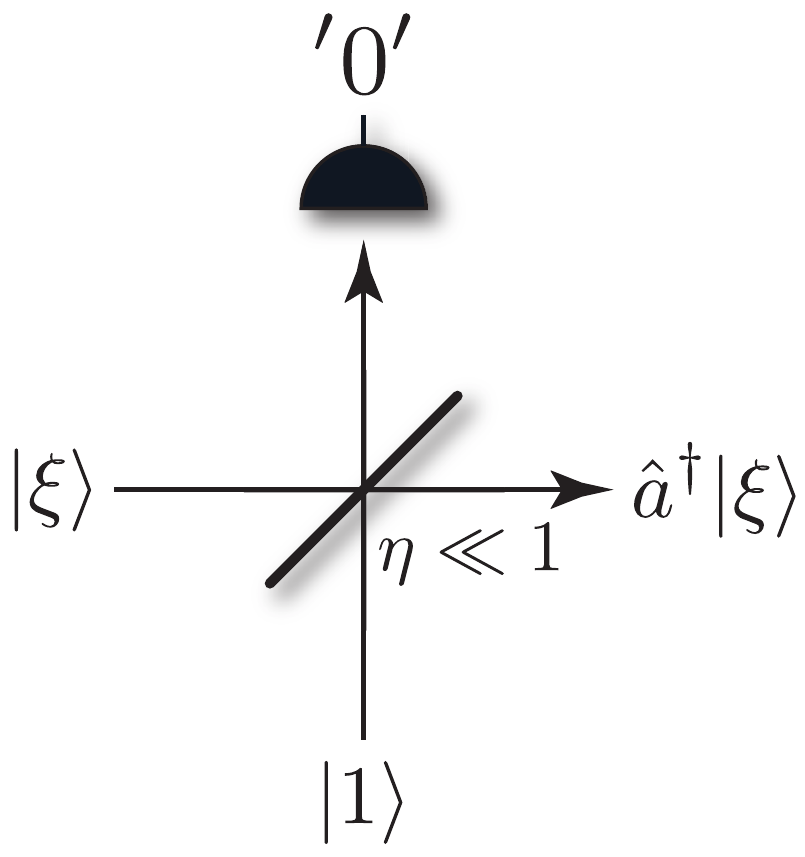}
\caption{Preparation of a photon-added SV state. A SV state is mixed with a single-photon state on a low reflectivity beamsplitter. The reflected mode is detected, and upon measuring the vacuum state we herald the preparation of the photon-added SV state in the other mode. The process is highly non-deterministic, but can be performed offline in advance.} \label{fig:prep}
\end{figure}

Photon-added SV states may be prepared by mixing a SV state (obtained from a degenerate parametric down-converter) with a single-photon state on a low reflectivity beamsplitter and post-selecting upon detecting the vacuum state in the reflected mode. Successful post-selection heralds the preparation of the photon-added SV state in the other mode. Thus, the preparation scheme is non-deterministic, but may be performed offline via trial-and-error in advance, enabling efficient state preparation. The preparation scheme is shown in Fig. \ref{fig:prep}.  Photon-subtracted SV states may be prepared similarly by sending in a squeezed state and a vacuum state to the inputs and post-selecting on one photon in the reflected mode. 

\textbf{Evolution:}
In both models, the input state is fed into a passive linear optics interferometer consisting of beamsplitters and phaseshifters, which in general transforms the creation operators according to a linear map,
\begin{equation} \label{eq:unitary_map}
\hat{U}\hat{a}_i^\dag\hat{U}^\dag \to \sum_j U_{i,j} \hat{a}_j^\dag,
\end{equation}
where $\hat{U}$ is an {$m\times m$} matrix.  For AA boson sampling, $\hat{U}_{AA}$ is chosen to be a Haar-random, unitary matrix.

Unlike the Fock state model, for PASSV boson sampling we consider an interferometer consisting of \emph{real} beamsplitters which implements an orthogonal matrix (also chosen to be Haar-random). Thus, for Fock state boson sampling \mbox{$\hat{U}_\mathrm{AA}\in SU(m)$}, whereas for PASSV boson sampling \mbox{$\hat{U}_\mathrm{SV} \in SO(m)$}. Reck \emph{et al.} showed that for both cases, any \mbox{$m\times m$} unitary or orthogonal matrix can be implemented with at most $O(m^2)$ optical elements, and an efficient algorithm for finding the decomposition exists \cite{bib:Reck94}.

It is important to discuss the complexity of choosing an orthogonal matrix instead of a unitary because one should be concerned with the possibility of choosing a subset of matrices from $SU(m)$, whose permanent is efficiently simulatable by a classical computer. If this were the case, the result would not be interesting, since the novelty of boson sampling is that it simulates a system which is classically intractable.  We will later prove this is not the case and that, in fact, the associated complexities are equivalent. 

\textbf{Output:}
The output state for the Fock state model after passing through the interferometer is thus,
\begin{eqnarray} \label{eq:aaoutput1}
\ket{\psi}_\mathrm{out}^\mathrm{AA} &=& \hat{U}_\mathrm{AA}\ket{\psi}_\mathrm{in}^\mathrm{AA} \nonumber \\
&=& \hat{U}_\mathrm{AA}\left[\hat{a}_1^\dag\dots\hat{a}_n^\dag\ket{0_1,\dots,0_m}\right] \nonumber \\
&=& \left[\hat{U}_\mathrm{AA}(\hat{a}_1^\dag\dots\hat{a}_n^\dag)\hat{U}_\mathrm{AA}^\dag\right]\hat{U}_\mathrm{AA}\ket{0_1,\dots,0_m} \nonumber \\
&=& \left[\hat{U}_\mathrm{AA}(\hat{a}_1^\dag\dots\hat{a}_n^\dag)\hat{U}_\mathrm{AA}^\dag\right]\ket{0_1,\dots,0_m},
\end{eqnarray}
where the last equality holds because $U_{AA}\ket{0}=\ket{0}$, i.e. $U_{AA}$ represents passive optics elements and hence cannot generate new photons. Since the unitary transforms the creation operators according to Eq. \ref{eq:unitary_map}, the output of the interferometer can also be represented as,
\begin{equation} \label{eq:aaoutput2}
\ket{\psi}_\mathrm{out}^\mathrm{AA}=\sum_{S}\gamma_S \ket{S_1,\dots,S_m},
\end{equation}
where $S$ is an output configuration of the $n$ photons with $S_i$ photons in the $i$th mode, and $\gamma_S$ is the corresponding amplitude. Note that \mbox{$\gamma_S \propto \text{Per}(U_{S})$}, where $U_{S}$ is an \mbox{$n\times n$} sub-matrix of $\hat{U}_{AA}$ given as a function of the configuration $S$. The number of distinct configurations is
\begin{equation} \label{eq:config}
|S| = \binom{n+m-1}{n},
\end{equation}
which can be easily verified to be the number of ways to configure $n$ indistinguishable photons into $m$ distinct modes. This expression grows superexponentially with $n$ from the earlier requirement that $m=\Omega(n^2)$.

\begin{figure}[t]
\includegraphics[width=0.65\columnwidth]{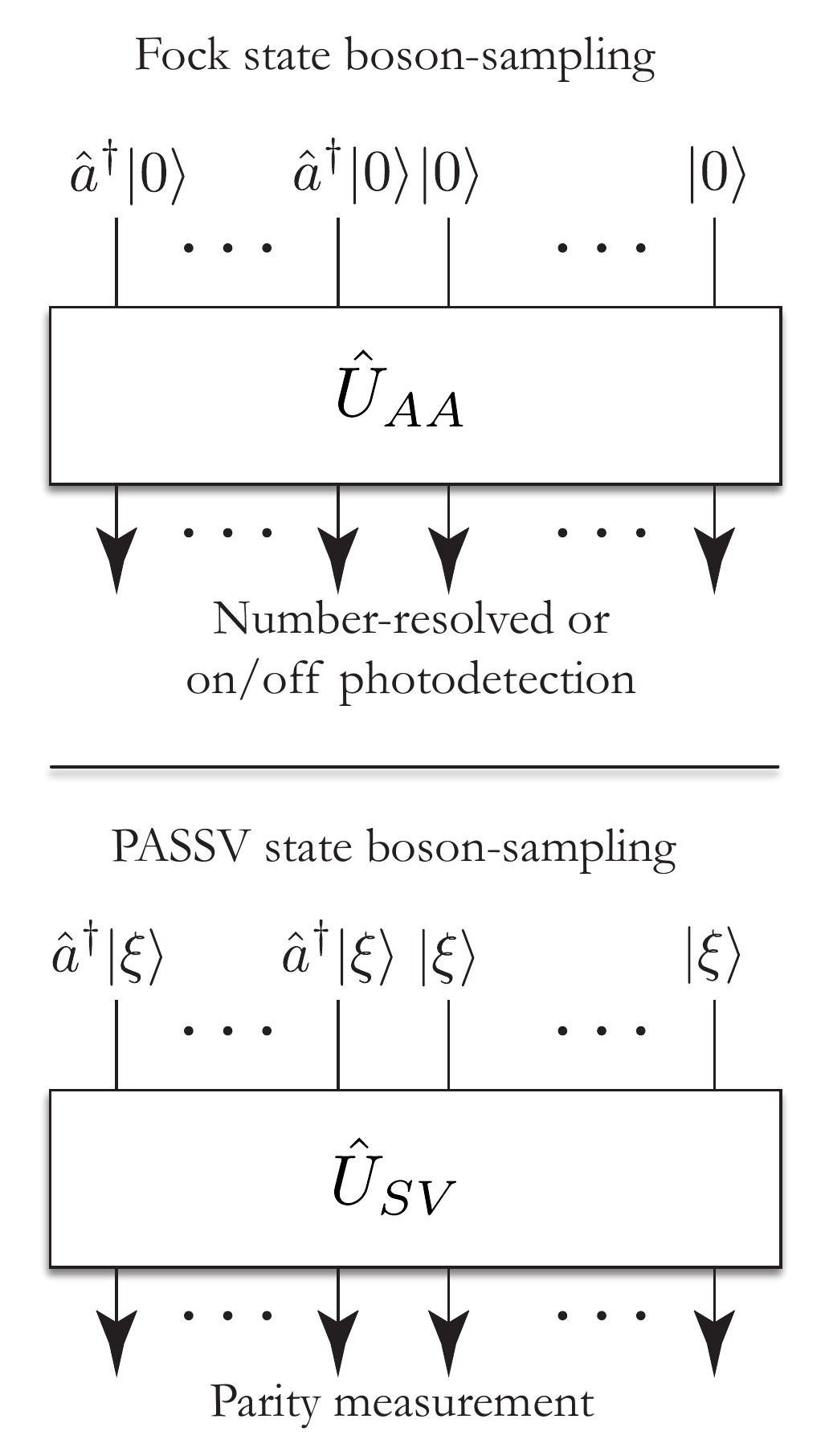}
\caption{(top) Fock state boson sampling. We feed an $m$-mode linear optics interferometer with $n$ single photons and \mbox{$m-n$} vacuum states. Following evolution, the state is sampled via coincidence number-resolved photodetection. (bottom) PASSV boson sampling. We prepare $n$ PASSV states and \mbox{$m-n$} SV states. Following evolution we perform coincidence parity measurement.}
\end{figure}

For PASSV boson sampling, we can use the same technique as in Eq. \ref{eq:aaoutput1}, such that the output state is,
\begin{eqnarray} \label{eq:svoutput1}
\ket{\psi}_\mathrm{out}^\mathrm{SV} &=& \hat{U}_\mathrm{SV}^\mathrm{}\ket{\psi}_\mathrm{in}^\mathrm{SV} \nonumber \\
&=& \left[\hat{U}_\mathrm{SV}(\hat{a}^\dag_1\dots\hat{a}^\dag_n)\hat{U}_\mathrm{SV}^\dag\right]\hat{U}_\mathrm{SV}\ket{\xi_1,\dots,\xi_m}. 
\end{eqnarray}
It was shown by Jiang \emph{et. al} \cite{bib:PhysRevA.88.044301} that for a pure product state input to a linear optical network, the output is entangled unless the input is either a tensor product of coherent states or a tensor product of squeezed states (with the same squeezing), provided that the network does not mix the squeezed and anti-squeezed quadratures. The latter condition is equivalent to the network comprising real beamsplitters. This condition is satisfied since \mbox{$\hat{U}_{SV}\in SO(m)$} and thus,
\begin{equation}
\label{eq:svoutput1b}
\ket{\psi}_\mathrm{out}^\mathrm{SV}= \left[\hat{U}_\mathrm{SV}(\hat{a}^\dag_1\dots\hat{a}^\dag_n)\hat{U}_\mathrm{SV}^\dag\right]\ket{\xi_1,\dots,\xi_m}.
\end{equation}
The leading operator corresponds to a configuration of $n$ creation operators as in Eq. \ref{eq:aaoutput1}. The output for a photon-added SV state input is therefore of the form,
\begin{equation} \label{eq:svoutput2}
\ket{\psi}_\mathrm{out}^\mathrm{SV} =\sum_{S}\gamma_{S}' \left[(\hat{a}_1^\dag)^{S_1} \dots (\hat{a}_m^\dag)^{S_m}\right]\ket{\xi_1,\dots,\xi_m},
\end{equation}
where,
\begin{equation}
\gamma_S' = \frac{\gamma_S}{\sqrt{S_1!\dots S_m!}}=\frac{\textrm{Per}(U_S)}{\sqrt{S_1!\dots S_m!}},
\end{equation}
but in the binary regime \mbox{$\gamma_S'=\gamma_S$}. Recall from Eq. \ref{eq:sv} that squeezed states represented in the Fock basis have only even photon-number terms. Thus, for a configuration $S$ where mode $i$ does not have a creation/annihilation operator acting on it, mode $i$ is a superposition of only even photon number states, whereas if $S$ applies a creation/annihilation operator to mode $i$ it contains only odd photon-number terms.

For photon-subtracted SV states the output is of the same form, replacing $\hat{a}^\dag_i$ with $\hat{a}_i$, but $\gamma_S$ will now relate to $\hat{U}^\dag_{SV}$ instead of $\hat{U}_{SV}$, which is also Haar-random, and thus has the same computational complexity.  We exclude the case of the photon-subtracted states when $\xi=0$ since $\hat{a}\ket{0}=0$.

\textbf{Measurement:}
The last step of boson sampling is to measure the output distribution. For Fock state boson sampling, this may be implemented via number-resolved photodetection. However, since \mbox{$m=\Omega(n^2)$}, \mbox{$S_i=\{0,1\}\,\,\forall \,\,i$} in Eq. \ref{eq:aaoutput2}, on/off (or `bucket') detectors are sufficient to recover the configuration $S$. Repeating the sampling procedure multiple times yields partial information of the joint photon-number distribution \mbox{$P_S=|\gamma_S|^2$}, which was shown by AA to be a computationally complex sampling problem.

For PASSV boson sampling, we perform a parity measurement capable of distinguishing only between odd and even photon-number. Such measurements are characterised by the measurement operators,
\begin{eqnarray}
\hat{\Pi}_+ &=& \ket{0}\bra{0} + \ket{2}\bra{2} + \ket{4}\bra{4} + \dots \\ \nonumber
\hat{\Pi}_- &=& \ket{1}\bra{1} + \ket{3}\bra{3} + \ket{5}\bra{5} + \dots
\end{eqnarray}
Most simply, one could implement this measurement using photon-number-resolving detectors. Measuring an even photon-number at output mode $i$ then implies that there was no creation/annihilation operator associated with that mode, whereas measuring an odd photon-number implies that there was.  This measurement thus perfectly recovers the configuration $S$, and hence continued sampling yields the desired distribution.  Since the squeezing parameter $\xi$ has no effect on the parity of the state, the sampling amplitudes are completely independent of the squeezing.  

More formally, in standard boson sampling we are sampling from a set of strings $s_i=\{s_i^{(1)},\dots,s_i^{(m)}\}$, where $s_i^{(j)}$ is the sampled photon-number in the $j$th mode associated with string $i$, of which there are an exponential number. In the limit of large $m$, $s_i^{(j)}\in\{0,1\}$. On the other hand, with PASSV boson sampling we are sampling from the same set of strings, with the same probability distribution, where now $s_i^{(j)}\in\{-1,1\}$.  This proves that PASSV boson sampling implements the same logical sampling problem as Fock state boson sampling, independent of the squeezing parameter.

\subsection{Complexity Concerns}
We previously mentioned, while discussing the evolution of the input state, whether choosing an orthogonal matrix has any implications for the complexity of PASSV sampling.  Since we have now shown that the PASSV model samples permanents of submatrices in the same way as Fock state sampling, this is the only barrier to completing our proof that the two models are in the same complexity class.

The first consideration is whether or not a Haar-random matrix in $SO(m)$ might have an efficiently computable exact or approximate permanent. The exact permanent case is known to be \#\textbf{P}-complete even for binary entries, \mbox{$U_{i,j}\in\{0,1\}$} \cite{bib:Valiant79}. There is also a known algorithm for efficiently approximating a permanent if the matrix has entries consisting of only non-negative real numbers. In the same work, it is shown that for a matrix with even a single negative entry, an efficient approximation algorithm would allow one to compute an \textit{exact} \mbox{$\{0,1\}$}-permanent efficiently \cite{bib:Jerrum04}. Although having to compute a difficult permanent is a necessary but not sufficient condition for computational hardness, since $SO(m)$ is considered to be universal for linear optics \cite{bib:Bouland}, there is no such complexity gap between unitary and orthogonal matrices.

More concretely, it has been shown that $SU(m)\subset SO(2m)$ \cite{bib:Georgi99}, i.e. for a $2m$-mode interferometer, the set of all orthogonal transformations includes all unitary $m$-mode transformations as a subgroup.  Thus, the complexity of sampling the output from a boson sampling device implementing an arbitrary matrix from $SO(2m)$ is at least as hard as sampling matrices from $SU(m)$, and for only a linear cost in the number of modes. Since trivially $SO(2m)\subset SO(2m+1)$, the same complexity extends to an odd number of modes as well.  Note that this also carries the implication that Fock state boson sampling itself remains hard under orthogonal transformations.

We can now conclude that PASSV boson sampling is in the same complexity class as the Fock state boson sampling proposed by AA.  Suppose that $A$ is some complexity class containing Fock state boson sampling.  Since the output of PASSV boson sampling is completely independent of the squeezing parameter $\xi$, we may assume without loss of generality that $\xi=0$.  In this limit, however, $\ket{\xi_i}=\ket{0_i}$ and thus, by construction, any instance of PASSV boson sampling reduces to an instance of Fock state boson sampling since $SO(m)\subset SU(m)$.  Thus, the class $A$ also contains PASSV boson sampling.  Conversely, suppose $B$ is some complexity class containing PASSV boson sampling.  Again choosing $\xi=0$, the inclusion  $SU(m)\subset SO(2m)$ similarly implies $B$ also contains Fock state boson sampling.

\section{Discussion}
Our result can be distilled to a relatively simple idea which is most evident in light of Eq. \ref{eq:aaoutput1}, where the ket acts as a `background' signal whose form is invariant under the evolution of $\hat{U}_{SV}$. Since the leading operator in Eq. \ref{eq:svoutput1b} takes exactly the same form as Eq. \ref{eq:aaoutput1}, we would like the ket to also be independent of the choice of $\hat{U}_{SV}$ under \emph{some} measurement, while still being distinguishable from a state which has an added or subtracted photon. It may be possible to use the same technique to characterize other states which implement a logically equivalent classically intractable sampling problem. A desirable goal would be to prove an even more experimentally friendly set of states and measurements that implements the same problem.

One criticism of PASSV boson sampling is that the use of photon-number resolving detectors to implement the parity measurement is experimentally harder than on/off detection. Whilst this is true, one does not need to distinguish between \emph{arbitrarily} large even and odd photon-number Fock states. For any given $\xi$ and error rate, one can truncate the maximum number of necessarily distinguishable Fock states. Indeed, PASSV boson sampling can be regarded as a generalization of Fock state boson sampling, since in the limit of small squeezing (\mbox{$\xi\rightarrow 0$}), the SV reduces to a vacuum state and an on/off detector suffices. For large squeezing, additional experimental hurdles may arise in reducing squeezing parameter error and in the increased sensitivity of squeezed states to noise.  We do not address these issues here.  Rather, despite PASSV states being more difficult to experimentally prepare, our goal is to theoretically demonstrate the non-uniqueness of Fock states for computationally hard sampling problems.

After having spent some effort showing that orthogonal matrices are sufficiently complex for PASSV sampling, a natural question is whether or not choosing a unitary matrix could change the complexity of the sampling problem.  Because Eq. \ref{eq:svoutput1b} no longer holds, we cannot establish a straightforward relationship between the output probabilities and submatrix permanents.  Conventional wisdom seems to suggest that the problem would not become easier.  In the limit of zero squeezing, we know there is no complexity divide because PASSV sampling reduces to Fock state sampling.  Thus, if a complexity divide did exist, then we would expect a complexity phase transition at $\xi=0$.  It may be possible to construct a more complicated measurement scheme which produces the same sampling probabilities.

We have shown a direct mapping between Fock state boson sampling and PASSV boson sampling. An open question in the field is `what classes of quantum states of light yield hard sampling problems with linear optics?' This result, in conjunction with previous results on photon-added coherent states and generalized cat states, demonstrates that there is a large class of non-Fock states, which yield sampling problems of equal computational complexity.

Importantly, unlike previous work on non-Fock state boson sampling, PASSV boson sampling operates in \emph{all} parameter regimes. Thus there are no bounds on the amount of squeezing and no approximations are made.

Whilst PASSV boson sampling may be experimentally more challenging than Fock state boson sampling, this result certainly confirms that there is nothing unique about the computational complexity of Fock states. In fact, there is a plethora of other quantum states exhibiting similar sampling complexity, and computational complexity appears to be a ubiquitous property of sampling quantum states of light.

We hope that future research will enable us to fully characterize what it is that makes a quantum optical system computationally hard, and what classes of states are required for computational complexity.

\begin{acknowledgments}
KRM and PPR would like to acknowledge the Australian Research Council Centre of Excellence for Engineered Quantum Systems (Project number CE110001013). JPD would like to acknowledge the Air Force Office of Scientific Research for support and both KPS and JPD would like to acknowledge the Army Research Office for support.  The authors would also like to acknowledge helpful discussions with Giulia Ferrini and Gavin Brennen.
\end{acknowledgments}

\bibliography{pssvbib}

\end{document}